\documentclass[10pt,journal,compsoc]{IEEEtran}

\usepackage{hyperref}       
\usepackage{url}            
\usepackage{amsfonts}       
\usepackage{microtype}      
\usepackage{xcolor}         
\usepackage{algorithm}      
\usepackage{algorithmic}
\usepackage{amsmath}
\usepackage{xcolor}
\usepackage{graphicx}

\usepackage{multirow}
\usepackage{multicol}
\usepackage{bbding}         
\usepackage{makecell}
\usepackage{tabularx}       
\usepackage{amsmath,amsfonts}
\usepackage{algorithmic}
\usepackage{algorithm}
\usepackage{array}
\usepackage[caption=false,font=normalsize,labelfont=sf,textfont=sf]{subfig}
\usepackage{url}
\usepackage{verbatim}
\usepackage{graphicx}
\usepackage{graphicx}  
\usepackage{subcaption}
\usepackage{booktabs}
\usepackage{multirow}

\newcommand{\bt}[1]{\textbf{#1}}

\begin{document}

\title{UniMoE-Audio: Unified Speech and Music Generation with Dynamic-Capacity MoE}

\author{Zhenyu Liu, Yunxin Li, Xuanyu Zhang, Qixun Teng, Shenyuan Jiang, Xinyu Chen, Haoyuan Shi, Jinchao Li, Qi Wang, Haolan Chen, Fanbo Meng, Mingjun Zhao, Yu Xu, Yancheng He, Baotian Hu, Min Zhang
\IEEEcompsocitemizethanks{
\IEEEcompsocthanksitem Zhenyu Liu, Yunxin Li, Xuanyu Zhang, Qixun Teng, Shenyuan Jiang, Xinyu Chen, Haoyuan Shi, Jinchao Li, Qi Wang, Baotian Hu and Min Zhang are with the Department of Computer Science and Technology, Harbin Institute of Technology, Shenzhen, China. (e-mail: liuzhenyuhit@gmail.com, hubaotian@hit.edu.cn, and zhangmin2021@hit.edu.cn)
\IEEEcompsocthanksitem Zhenyu Liu, Baotian Hu and Min Zhang are also with the Shenzhen Loop Area Institute, Shenzhen, China.
\IEEEcompsocthanksitem Baotian Hu is the corresponding author. (e-mail: hubaotian@hit.edu.cn)
}}



\markboth{Journal of \LaTeX\ Class Files,~Vol.~14, No.~8, August~2021}%
{Shell \MakeLowercase{\textit{et al.}}: A Sample Article Using IEEEtran.cls for IEEE Journals}


\IEEEtitleabstractindextext{%
\begin{abstract}
Recent advances in unified multimodal models indicate a clear trend towards comprehensive content generation. However, the auditory domain remains a significant challenge, with music and speech often developed in isolation, hindering progress towards universal audio synthesis. This separation stems from inherent task conflicts and severe data imbalances, which impede the development of a truly unified audio generation model. 
To address this challenge, we propose UniMoE-Audio, a unified speech and music generation model within a novel Dynamic-Capacity Mixture-of-Experts (MoE) framework. Architecturally, UniMoE-Audio introduces a Top-P routing strategy for dynamic expert number allocation, and a hybrid expert design comprising routed experts for domain-specific knowledge, shared experts for domain-agnostic features, and null experts for adaptive computation skipping. 
To tackle data imbalance, we introduce a three-stage training curriculum: 1) Independent Specialist Training leverages original datasets to instill domain-specific knowledge into each “proto-expert” without interference; 2) MoE Integration and Warmup incorporates these specialists into the UniMoE-Audio architecture, warming up the gate module and shared expert using a subset of balanced dataset; and 3) Synergistic Joint Training trains the entire model end-to-end on the fully balanced dataset, fostering enhanced cross-domain synergy. Extensive experiments show that UniMoE-Audio not only achieves state-of-the-art performance on major speech and music generation benchmarks, but also demonstrates superior synergistic learning, mitigating the performance degradation typically seen in naive joint training. Our findings highlight the substantial potential of specialized MoE architecture and curated training strategies in advancing the field of universal audio generation. Homepage: \url{https://mukioxun.github.io/Uni-MoE-site/home.html}.
\end{abstract}

}

\maketitle

\IEEEdisplaynontitleabstractindextext



\IEEEpeerreviewmaketitle

\section{Introduction}

\begin{figure*}[htb!]
    \centering
    \includegraphics[width=\linewidth]{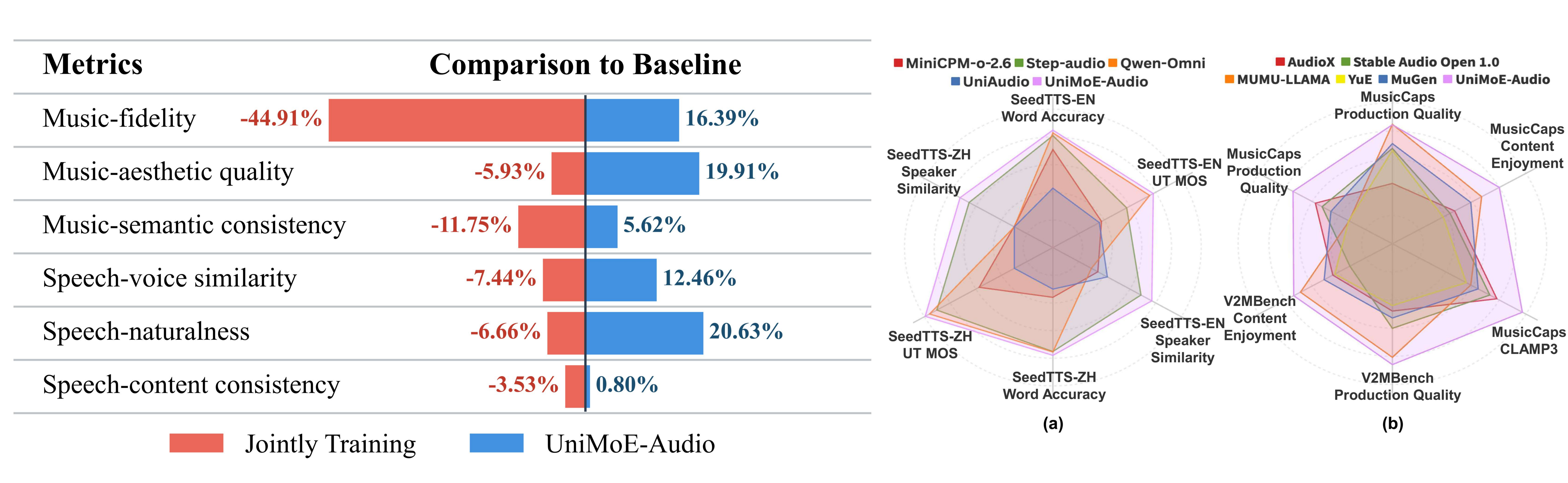}
    \vspace{-5mm}
    \caption{Performance of UniMoE-Audio.  \textbf{Left:} Comparison against specialized baselines reveals the failure of naive joint training, which causes a clear performance degradation on speech generation and more significant decline on music generation. In contrast, our UniMoE-Audio yields synergistic gains across both tasks. \textbf{Right:} Radar charts show UniMoE-Audio achieving the best comprehensive performance against leading models on a wide array of speech (a) and music (b) metrics.}
    \label{fig:motivation}
\end{figure*}

\IEEEPARstart{A} hallmark of human intelligence is the seamless ability to perceive, reason, and create across multiple modalities, effortlessly blending language, vision, and audio. Emulating this holistic capability represents a grand challenge and a core objective in the pursuit of more general artificial intelligence. The recent ascendancy of Large Language Models (LLMs) has served as a powerful catalyst, paving the way for unified models that can understand and generate content across these diverse data streams. Significant progress has been made in systems that jointly process text, images, video, and even speech within a single architecture~\cite{DBLP:conf/acl/ZhanDYZZLZYZL0F24,wu2025qwen,DBLP:journals/corr/abs-2503-20215,DBLP:journals/corr/abs-2506-09344,DBLP:journals/corr/abs-2504-18425,DBLP:journals/corr/abs-2502-11946}. Nevertheless, a critical imbalance persists in the treatment of the auditory domain. While speech has been a primary focus of integration~\cite{DBLP:journals/corr/abs-2504-18425,DBLP:journals/corr/abs-2502-11946}, music—a domain of comparable complexity and cultural richness—remains largely siloed and excluded from these unified frameworks. This fundamental omission not only curtails the ambition of universal audio synthesis but also stands as a significant impediment to developing AI with truly comprehensive multimodal intelligence.

The primary obstacle to unifying speech and music generation stems from two fundamental challenges. The first is \textbf{task conflict}, arising from the divergent objectives of speech and music generation. The former is primarily concerned with semantic intelligibility and speaker identity, whereas the latter focuses on capturing complex structures like harmony and rhythm. This divergence creates conflicting optimization pressures within a shared model, where progress on one task can impede the other. Recently, the MoE paradigm has emerged as a promising architecture for mitigating conflicts of multimodal understanding~\cite{DBLP:journals/corr/abs-2401-15947,DBLP:journals/pami/LiJHWZLMZ25,DBLP:journals/corr/abs-2506-09344}. Despite these advances, its application and further optimization for unified audio generation remain largely unexplored.
Beyond task conflict, another major hurdle is \textbf{data imbalance}. High-quality, large-scale speech corpora are far more abundant than their musical counterparts. The detrimental effects of this disparity are evident in prior work~\cite{DBLP:conf/icml/YangT0HLGCSZ0ZW24}. Consequently, a naive joint training approach often allows the data-rich speech task to dominate the learning process, resulting in a substantial degradation in musical quality. Our preliminary experiments empirically confirm this degradation (Figure~\ref{fig:motivation}), showing that a jointly trained model performs significantly worse than specialized models, with the performance drop being particularly severe for the data-scarce music task. Therefore, the central scientific question we address is: \textit{\textbf{how to overcome both task conflict and data imbalance, enabling a shared model to master speech and music generation synergistically?}}

Our approach addresses these challenges at both the architectural and training curriculum levels. Architecturally, we propose UniMoE-Audio, which leverages a novel Dynamic-Capacity MoE for mitigating task conflict. Instead of directly applying the conventional MoE, we provide two key architectural optimizations to improve both routing flexibility and functional decoupling. 
First, we introduce a dynamic-capacity routing strategy that replaces the conventional fixed-capacity routing. Based on the Top-P sampling, this strategy dynamically adjusts the number of experts allocated to each token based on their complexity, thus enabling more flexible expert combinations. Second, we present a hybrid expert design to establish clear functional specialization, comprising: 1) conditional routed experts for domain-specific knowledge; 2) constantly active shared experts to handle domain-agnostic features; and 3) null experts to skip computation adaptively.

While our architecture provides the structural means to mitigate task conflict, we introduce a tightly coupled three-stage training curriculum to address data imbalance. The curriculum unfolds as follows: (1) \textbf{Independent Specialist Training} leverages the original, uncurated datasets to instill domain-specific knowledge into each ``proto-expert" without interference. (2) \textbf{MoE Integration and Warmup} then integrates these specialists into the UniMoE-Audio architecture. This stage begins by creating a curated, balanced dataset via a rigorous data filtering pipeline. To ensure training stability, the newly added components (i.e. the gate module and the shared expert) are then exclusively warmed up on a small subset of this curated data. (3) \textbf{Synergistic Joint Training} finally trains the entire model on the full balanced dataset, fostering effective knowledge transfer across domains.
 
Our main contributions can be summarized as follows:
\begin{itemize}
\item We propose UniMoE-Audio, a unified speech and music generation model based on a novel Dynamic-Capacity MoE framework. By integrating a Top-P routing strategy for adaptive resource allocation and a hybrid expert design for functional decoupling, our architecture effectively mitigates the inherent task conflict between speech and music generation.

\item  To leverage this architecture and tackle data imbalance, we introduce a data-aware, three-stage training curriculum. This curriculum systematically overcomes the data imbalance challenge by orchestrating independent specialist training, router warmup, and synergistic joint training, enabling robust and effective learning from highly imbalanced data sources without resorting to conventional data sampling.

\item We provide extensive experiments to show the UniMoE-Audio's effectiveness, achieving state-of-the-art or competitive performance on major speech and music generation benchmarks. Furthermore, our in-depth analysis reveals the dynamic activation patterns of the MoE model, offering valuable insights into how the unified MoE model navigates diverse audio generation tasks.

\end{itemize}

\section{Related Work}

\subsection{Domain-Specific Audio Generation Models}

\noindent\textbf{Large Spoken Models}. The paradigm of generative AI, powered by Large Language Models (LLMs), has recently catalyzed a revolution in text-to-speech (TTS), giving rise to the field of the Large Spoken Models. This approach fundamentally reframes speech synthesis as a conditional language modeling problem. Typically, a Speech LLM consists of a large, decoder-only Transformer and a neural audio codec. Given a textual prompt and optional voice conditions, the Transformer autoregressively generates a sequence of discrete audio tokens, which are then converted back into a continuous waveform by the codec. This framework has enabled unprecedented capabilities in zero-shot voice cloning and expressive, controllable speech generation.
The seminal work in this area, VALL-E~\cite{DBLP:journals/corr/abs-2406-05370}, pioneered this approach by discretizing speech into acoustic tokens via the EnCodec~\cite{DBLP:journals/corr/abs-2210-13438} and modeling them conditioned on text. This breakthrough laid the groundwork for a proliferation of subsequent models, including VALL-E X~\cite{DBLP:journals/corr/abs-2303-03926}, SpearTTS~\cite{DBLP:journals/tacl/KharitonovVBMGP23}, and Make-a-Voice~\cite{DBLP:journals/corr/abs-2305-19269}, which further refined tokenization schemes and text-to-acoustic alignment. Building on this foundation, the field has seen rapid advancements towards greater robustness and versatility. For instance, CosyVoice~\cite{DBLP:journals/corr/abs-2407-05407} leverages a multi-task, multi-stage training curriculum to achieve state-of-the-art performance across a wide array of speech synthesis tasks. Concurrently, StepAudio~\cite{DBLP:journals/corr/abs-2502-11946} demonstrates the power of training on massive-scale synthetic data to produce exceptionally high-fidelity speech with rich emotional and stylistic diversity.

\noindent\textbf{Large Music Models}. Mirroring the evolution in speech synthesis, the field of music generation has also increasingly adopted the Large Language Model paradigm, reframing music composition as a sequence generation task guided by textual or visual prompts. While diffusion-based models like MusicLM~\cite{DBLP:journals/corr/abs-2301-11325} and Stable Audio Open~\cite{DBLP:conf/icassp/EvansPCZTP25} have achieved remarkable results, autoregressive models have demonstrated a compelling alternative. MusicGen~\cite{DBLP:conf/nips/CopetKGRKSAD23} was a pivotal work that validated the feasibility of modeling music with a single Transformer decoder, generating high-fidelity music from discrete tokens. Pushing the boundaries further, subsequent works have explored more complex architectures and functionalities. Built upon the architecture of Llama2~\cite{DBLP:journals/corr/abs-2307-09288}, YuE~\cite{DBLP:journals/corr/abs-2503-08638} introduced a track-decoupled prediction strategy to handle long-form music generation. MuMuLlama~\cite{DBLP:journals/corr/abs-2412-06660} introduces multimodal music generation by jointly training on text-to-music and vision-to-music tasks. These advancements collectively indicate the power and viability of autoregressive framework for controllable music synthesis.

While the aforementioned studies demonstrate substantial advancements in speech and music generation, they primarily focus on advancing the state-of-the-art within their respective domains. Our work, in contrast, shifts the focus from domain-specific excellence to the challenge of cross-domain unification. This line of inquiry is prompted by the observation that both fields, despite their distinct objectives, have independently converged on a similar technical paradigm: autoregressive modeling of discrete audio tokens. This parallel evolution suggests the potential for a single unified architecture that handle both speech and music generation, yet the feasibility and inherent complexities of such unification remain largely unexplored. Therefore, our work represents a foundational investigation into this underexplored area, aiming to broaden the scope of what generative audio models can achieve.

\subsection{Unified Audio Generation Models}

The ambition of a universal audio model has prompted several initial investigations into unifying diverse audio generation tasks within a single framework. A notable early attempt, UniAudio~\cite{DBLP:conf/icml/YangT0HLGCSZ0ZW24}, proposed a general-purpose text-to-audio model capable of generating various types of audio. However, as a naive joint training approach, it reportedly suffered from the problem of data imbalance, leading to limited performance on data-scarce tasks such as music generation. More recently, AudioX~\cite{DBLP:journals/corr/abs-2503-10522} demonstrated impressive capabilities in generating sound effects and music from multimodal inputs like text, images, and video, utilizing a Diffusion Transformer architecture. While powerful, its scope notably omits speech generation, a prevalent and critical audio modality, thus not addressing the full challenge of speech-music unification.
In contrast to these approaches, our work directly confronts the core challenges that have hindered previous unification efforts. Rather than relying on simple joint training, we propose a framework that explicitly accounts for the inherent differences between audio modalities. Specifically, we leverage a MoE architecture to mitigate task conflict and a data-aware, three-stage training curriculum to address data imbalance, aiming to provide a more principled and effective pathway toward truly unified and high-fidelity audio generation.

\section{UniMoE-Audio}

\begin{figure*}[ht!]
    \centering
    \includegraphics[width=\linewidth]{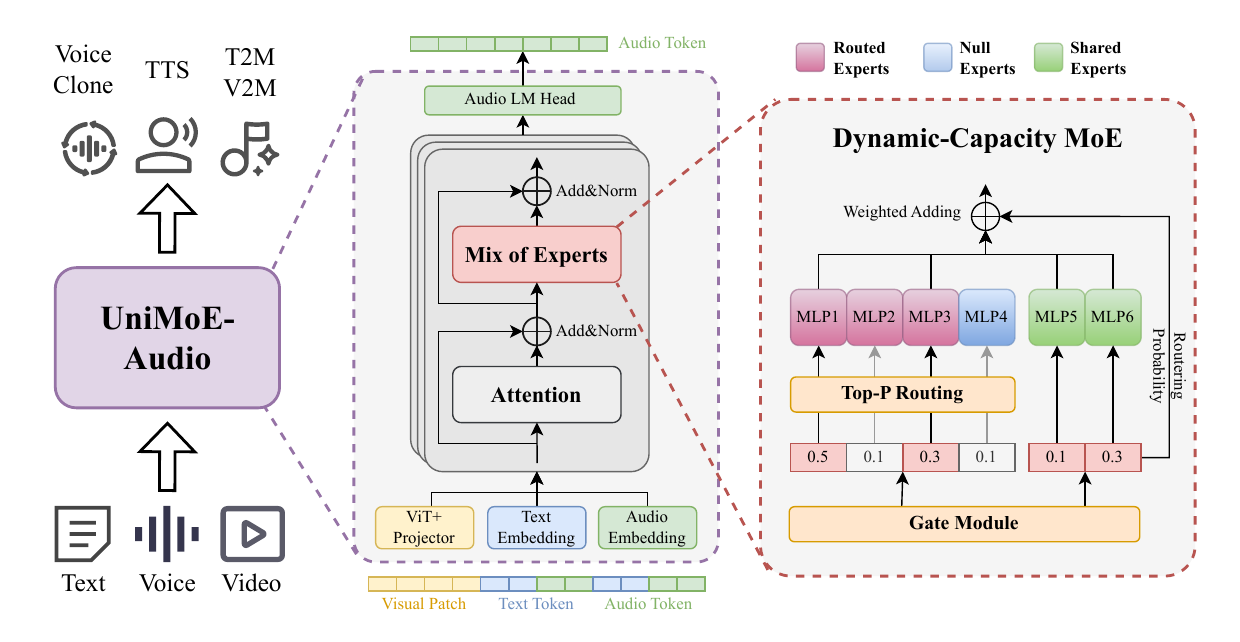}
    \vspace{-5mm}
    \caption{An overview of the UniMoE-Audio framework. \textbf{Left:} UniMoE-Audio is a unified model capable of performing speech and music generation by leveraging multimodal conditional inputs, including Voice Cloning, Text-to-Speech (TTS), Text-to-Music (T2M), and Video-to-Music (V2M). \textbf{Center:} The core architecture of our model is a Transformer with Dynamic-Capacity MoE layers. \textbf{Right:} We propose a novel Top-P routing strategy, which dynamically selects the number of experts allocated to each token based on their complexity.}
    \label{fig:architecture}
\end{figure*}

Our proposed model, UniMoE-Audio, is a unified generative framework designed to synthesize both speech and music from multimodal inputs, including text, audio, and video. As illustrated in Figure~\ref{fig:architecture}, the core innovation of the architecture lies in the Dynamic-Capacity MoE implementation, which deviates from conventional MoE in two aspects: (1) a novel Top-P routing strategy for dynamic experts number allocation, and (2) a hybrid expert design comprising routed, shared, and null experts.

\subsection{Input Representation and Tokenization}

\noindent\textbf{Audio Tokenization}. Following established practices in audio generation, we employ a neural audio codec to transform continuous waveforms into a sequence of discrete acoustic tokens. Specifically, we utilize the DAC codec~\cite{DBLP:conf/nips/KumarSLKK23}, which represents each audio frame using a multi-channel codebook. Unlike some works~\cite{DBLP:journals/corr/abs-2410-00037,DBLP:conf/icml/YangT0HLGCSZ0ZW24} that employ the Depth Transformer to predict tokens for each channel sequentially, we adopt a more parameter-efficient approach. Our model predicts all channels with a multi-head output layer. This design avoids the introduction of additional sequential modules, thereby reducing the overall parameter count and computational latency.

\noindent\textbf{Visual Embedding}. To process visual inputs (e.g., from video), we follow the Qwen-VL~\cite{DBLP:journals/corr/abs-2409-12191}, using a Visual Transformer (ViT) to encode the input image into patches. These visual features are then mapped into the language model's embedding space via a projector module, yielding a sequence of soft visual tokens that can be seamlessly integrated with text and audio representations.

\subsection{Dynamic-Capacity MoE}

A primary limitation of conventional MoE models is their static Top-K routing strategy, which allocates a fixed number of experts to each token. This approach is computationally sub-optimal, as it may over-allocate computational resources to simple tokens while under-powering complex ones that require more extensive processing. To address this, we introduce a Top-P routing mechanism that dynamically allocates the number of activated experts for each token based on the routing probability of the router module.

Given an input tensor $X \in \mathbb{R}^{N \times d}$ for an FFN layer, where $N$ is the sequence length and $d$ is the hidden dimension, a linear module first computes the gating probabilities for all $E$ experts:
\begin{equation}
P = \text{Softmax}(XW_g),
\end{equation}
where $W_g \in \mathbb{R}^{d \times E}$ is the trainable gating matrix and $P \in \mathbb{R}^{N \times E}$ represents the probability distribution over experts for each token.

We interpret this distribution $P$ as the router's confidence. The objective is to select the smallest set of experts whose cumulative probability exceeds a predefined threshold $p$, thereby balancing computational cost and predictive accuracy. This can be formulated as finding an index set $I$ for each token such that:
\begin{equation}
I = \arg\min_{I'} |I'| \quad \text{s.t.} \quad \sum_{i \in I'} P_i \ge p.
\end{equation}
To efficiently solve this, we employ the classic Top-P sampling algorithm, sorting expert probabilities in descending order and selecting the smallest set whose cumulative sum exceeds the threshold  $p$. The experts included in this sum are selected for computation. This approach naturally links the number of selected experts to the complexity of token, which is reflected in the router’s probability distribution: low-entropy distributions correspond to simpler tokens, while high-entropy ones indicate more complex tokens requiring more experts.

The final output of the MoE layer is a weighted sum of the outputs from the selected experts, where the weights are the normalized gating probabilities:
\begin{equation}
O = \sum_{i \in I} \frac{P_i}{\sum_{j \in I} P_j} E_i(X),
\end{equation}
where $I$ is the set of selected expert indices for a given token, and $E_i(X)$ is the output of the $i$-th expert.

While routed experts excel at learning domain-specific knowledge through conditional activation, they are inefficient for acquiring common knowledge, as inactive experts are excluded from the learning process. To address this, we functionally decouple the expert pool. Specifically, we incorporate a set of shared experts that operate in parallel with the routed ones, which is constantly activated for all tokens, aimed at capturing common knowledge and offloading the computational burden in routed experts, allowing the routed experts to dedicate their full capacity to mastering domain-specific patterns.

Furthermore, while our proposed routing strategy enables adaptive expert allocation, the range of activated expert number is inherently constrained. For a set of $N_{r}$ routed experts and the probability threshold $p$, the number of activated experts is confined to the range $[1, \lceil pN_{r} \rceil]$. This prevents true computation skipping for simple tokens or activating all the router experts for the most demanding ones, limiting the model's adaptive potential. To overcome this, we introduce the null expert: a parameter-free module whose output is a constant zero tensor. By incorporating $N_{n}$ null experts into the routing pool, the possible number of activated routed experts now spans the expanded range of $[0, \lceil p(N_{r} + N_{n}) \rceil]$. This not only enhances the combinatorial flexibility of expert selection but also enables true adaptive computation skipping.

\section{Training}

The successful unification of speech and music generation hinges not only on the model architecture but also on a training strategy that can effectively navigate the challenges of data imbalance and task conflict. To this end, we devise a comprehensive approach encompassing both rigorous data governance and a principled, three-stage training curriculum.

\subsection{Training Data}

\begin{table}[ht]
\centering
\scriptsize
\caption{Overview of Datasets Used in Different Tasks}
\label{tab:data}
\begin{tabular}{llll}
\toprule
\textbf{Task} & \textbf{Datasets} & \textbf{Number} & \textbf{Duration (hours)} \\
\midrule
\multirow{2}{*}{Speech Synthesis} & Mandarin TTS &180K & 20K \\
 & English TTS & 100K & 10K \\
\cmidrule(lr){1-4} 
\multirow{3}{*}{Text-to-Music} & Free-music-archive~\cite{DBLP:conf/ismir/DefferrardBVB17} & 106K & 8.2K \\
 & MusicNet~\cite{DBLP:conf/iclr/ThickstunHK17} & 320 & 37 \\
 & MU2Gen~\cite{DBLP:journals/corr/abs-2412-06660} & 22K & 1.2K \\
\cmidrule(lr){1-4} 
Video-to-Music & V2M~\cite{DBLP:conf/cvpr/Tian0YPL0CXG25} & 20K & 600 \\
\bottomrule
\end{tabular}
\end{table}

To support the unified generation of speech and music, we constructed a comprehensive, multi-task dataset collection, detailed in Table~\ref{tab:data}. Our data strategy involves two key components: a large-scale, imbalanced raw dataset for initial specialist training, and a smaller, high-quality balanced dataset for subsequent MoE joint training.

Our data curation process began with the collection of extensive raw data across four distinct categories: Chinese TTS (ZhTTS), English TTS (EnTTS), Text-to-Music (T2M), and Video-to-Music (V2M). As shown in Table~\ref{tab:data}, this resulted in a highly imbalanced corpus, with speech data (approx. 30K hours) vastly outnumbering music data (approx. 10K hours). All data underwent a rigorous pipeline of automatic annotation, multi-metric filtering, and deduplication to ensure quality. This large-scale raw dataset is exclusively used in the pre-training stage of our training curriculum to train the individual proto-experts, allowing each specialist to leverage the maximum available data for its domain without being affected by the data imbalance of other tasks.

To mitigate task dominance in the later joint training stages, we constructed a high-quality, balanced dataset. This dataset was created by carefully sampling 15K high-quality samples from each of the four task domains (ZhTTS, EnTTS, T2M, V2M) from our curated raw data pools. This results in a final balanced set of 60K samples, ensuring that the model receives equal exposure to each task during the critical MoE warmup and synergistic joint training stages. This balanced approach is crucial for preventing the model from developing a bias towards the data-rich speech tasks and for fostering true cross-domain synergy.

\subsection{Three-stage Training Curriculum}

A naive joint training approach on the imbalanced dataset would inevitably lead to the data-rich speech task dominating the learning process. Conversely, simple up-sampling or down-sampling from the outset either sacrifices data diversity or discards valuable resources. To systematically circumvent this dilemma, we propose a data-aware, three-stage training curriculum, designed to decouple task-specific learning from synergistic optimization.

\noindent\textbf{Independent Specialist Training}. The primary objective of this stage is to mitigate task conflict at its source and maximize data utilization. We leverage the full, imbalanced raw datasets to train separate, dense models for each task, as listed in Table~\ref{tab:setting}. This complete isolation allows each model—which will serve as a "proto-expert"—to master its domain-specific knowledge without interference from other tasks. This process effectively injects specialized knowledge into the parameters of each future expert, pre-assigning their intended function before they are integrated.

\noindent\textbf{MoE Integration and Warmup}. In this stage, we transition from individual specialists to the unified UniMoE-Audio model. Specifically, the Feed-Forward Network (FFN) block from each of the four proto-experts is split into two halves, creating a total of eight domain-specialized routed experts. Shared components, such as the attention and layer normalization modules, are initialized by averaging their corresponding parameters from all four proto-experts, while the vision transformer inherits its parameters directly from the ``Expert-V2M" model.

Once assembled, the weights of these pre-trained routed experts are initially frozen. The key challenge here is to stably integrate the newly introduced, randomly initialized components: the routing module and the two shared experts. Naive joint training would expose the well-trained experts to arbitrary routing decisions, risking catastrophic forgetting. To prevent this, we perform a crucial calibration step: using only the balanced dataset, we exclusively train the gate modules and shared experts. This allows the routers to learn meaningful dispatch patterns based on the experts' pre-trained specializations and stabilizes the shared components before full-model training.

\noindent\textbf{Synergistic Joint Training}. With a stable and calibrated routing mechanism in place, the final stage aims to foster synergistic learning across all tasks. We unfreeze the entire model and conduct end-to-end fine-tuning on the larger, balanced fine-tuning dataset. To maintain routing efficiency and prevent the collapse of expert specialization during joint training, we employ an auxiliary load-balancing loss. The weight of this loss is linearly annealed over the course of training. Initially, a high weight encourages the model to prioritize balanced expert utilization, promoting exploration. As training progresses, the weight decreases, shifting the optimization focus toward maximizing the primary sequence generation objective and exploiting the learned, efficient routing patterns for superior performance.

\section{Experiments}

\subsection{UniMoE-Audio Setting}

\begin{table*}[t]
\renewcommand\arraystretch{1.20}
\caption{Model configurations and parameters of all model variants used in our experiments. The Unify-Baseline and UniMoE-Audio models are designed to have a comparable total parameter count for a fair comparison.}
\label{tab:setting}
\centering
\begin{tabular}{l|cccc}
\hline
\textbf{Name} & \textbf{Task} & \textbf{Architecture} & \textbf{Activated Param} & \textbf{Total Param}\\

\hline
{Expert-ZhTTS} & Chinese TTS & Dense   & 3.1B & 3.1B \\
{Expert-EnTTS} & English TTS & Dense   & 3.1B & 3.1B \\
{Expert-T2M} & Text to Music & Dense   & 3.1B & 3.1B \\
{Expert-V2M} & Video to Music & Dense   & 3.1B & 3.1B \\
{Unify-Baseline} & Unify Audio Generation  & Dense  & 7.1B & 7.1B \\
{UniMoE-Audio} & Unify Audio Generation & Dynamic-Capacity MoE & Avg: 4.8B (Min: 2.8B, Max: 5.9B) & 7.1B \\
\hline

\end{tabular}
\end{table*}

This section outlines the configurations of all model variants evaluated in our experiments, with key specifications summarized in Table~\ref{tab:setting}. Our experiments involve three main categories of models, all developed based on the Qwen2.5VL architecture:

\begin{itemize}
    \item \textbf{Specialist Models:} Four 3.1B dense models, each trained on a single task (Chinese TTS, English TTS, T2M, V2M). These serve as the foundational "proto-experts" and represent the performance of dedicated, single-task systems.
    \item \textbf{Unify-Baseline:} A 7.1B dense model trained via direct joint training on the combined dataset. It serves as a strong baseline to ablate the benefits of our MoE architecture and specialized training curriculum.
    \item \textbf{UniMoE-Audio:} Our proposed 7.1B unified model, featuring a Dynamic-Capacity MoE architecture. Its activated parameter count is variable, governed by a Top-P routing strategy ($p=0.7$), averaging approximately 4.8B activated parameters during inference.
\end{itemize}

\subsection{Implementation Details}

We employ the AdamW~\cite{DBLP:conf/iclr/LoshchilovH19} optimizer in conjunction with a cosine learning rate scheduler across all training stages. Subsequently, in the independent specialist training stage, we utilize 48 Ascend 910B GPUs, with a global batch size of 48 and a base learning rate of 1e-4. In the MoE integration and warmup stage, we utilize 196 Ascend 910B GPUs for MoE training, with a global batch size of 784 and a base learning rate of 3e-5. Finally, in the synergistic joint training stage, we utilize 196 Ascend 910B GPUs, with a global batch size of 3136 and a base learning rate of 1e-5. We adopt expert parallelism with four-way partitioning, meaning only two routed experts are loaded on each GPU.

\subsection{Evaluation Setting}

Our evaluation setting comprehensively assesses both speech and music generation capabilities across a range of standard benchmarks and metrics.

\noindent\textbf{Speech Synthesis}. 
For speech synthesis, we evaluate models on both English and Mandarin benchmarks, focusing on three primary aspects: content intelligibility, speaker similarity, and perceptual quality. Our evaluation benchmark includes the Seed-TTS test set~\cite{DBLP:journals/corr/abs-2406-02430}, the LibriSpeech test-clean set~\cite{DBLP:conf/icassp/PanayotovCPK15}, and AISHELL-3~\cite{DBLP:conf/interspeech/ShiBXZL21}. For content intelligibility and perceptual quality, we utilize a consistent voice prompt to isolate the model's generative quality from prompt variations.
\begin{itemize} 
\item \textbf{Content Intelligibility} is measured by Word Error Rate (WER) for English and Character Error Rate (CER) for Mandarin, computed using the Whisper-large-v3~\cite{DBLP:conf/icml/RadfordKXBMS23} and Paraformer-zh\cite{DBLP:conf/interspeech/GaoZ0Y22} ASR engines, respectively. 
\item \textbf{Perceptual Quality} is assessed using UTMOS~\cite{DBLP:conf/interspeech/SaekiXNKTS22}, a neural MOS predictor that serves as an objective proxy for subjective human ratings.
\item \textbf{Speaker Similarity} is quantified by the cosine similarity of speaker embeddings extracted from a fine-tuned WavLM model, following the methodology of Seed-TTS~\cite{DBLP:journals/corr/abs-2406-02430}.
\end{itemize}

\noindent\textbf{Music Generation}. 
For music generation, we evaluate both text-to-music (T2M) and video-to-music (V2M) tasks, assessing semantic alignment, audio quality, and aesthetic quality. The T2M task is evaluated on MusicCaps~\cite{DBLP:journals/corr/abs-2301-11325} and V2M-bench\cite{DBLP:conf/cvpr/Tian0YPL0CXG25}, and the V2M task is evaluated on V2M-bench. Notably, to align with the setting of MusicCaps, all video and audio samples from V2M-Bench are segmented into 10-second clips.
\begin{itemize} 
\item \textbf{Semantic Alignment} between text and audio is measured using CLAP score~\cite{DBLP:conf/icassp/WuCZHBD23}. To provide a more robust assessment, we also report the CLaMP3 score~\cite{DBLP:conf/acl/WuGYJDXNLY025}, which leverages a more advanced multilingual framework. 
\item \textbf{Audio Quality and Diversity} are evaluated using a suite of metrics: Fréchet Audio Distance (FAD)~\cite{DBLP:conf/interspeech/KilgourZRS19} with OpenL3 embeddings, Kullback-Leibler (KL) divergence based on PaSST~\cite{DBLP:conf/interspeech/KoutiniSEW22} predictions, and Inception Score (IS). 
\item \textbf{Aesthetic Quality} is evaluated using three specialized metrics from~\cite{DBLP:journals/corr/abs-2502-05139}: Production Complexity (PC), Production Quality (PQ), and Content Enjoyment (CE). 
\end{itemize}

\subsection{Overall Performance}

\begin{table*}[ht]
\centering
\caption{Performance on English and Mandarin speech synthesis benchmarks. The best performance for each metric is highlighted in \textbf{bold}. WER and CER measure content intelligibility (lower is better), UTMOS measure perceptual quality (higher is better), and SIM measure speaker similarity with reference voice. UniMoE-Audio achieves state-of-the-art (SOTA) results on multiple key metrics and demonstrates highly competitive performance on others.
}
\label{tab:TTS_main}
\begin{tabular}{l|ccc|ccc|cc|cc}
\toprule

& \multicolumn{3}{c|}{\textbf{SeedTTS-EN}} & \multicolumn{3}{c|}{\textbf{SeedTTS-ZH}} &  \multicolumn{2}{c|}{\textbf{librispeech}}  &  \multicolumn{2}{c}{\textbf{AISHELL-3}} \\
\cmidrule(lr){2-11}

\multirow{-2}{*}{\textbf{Method}} & \textbf{WER↓} & \textbf{UTMOS ↑} & \textbf{SIM↑} & \textbf{CER↓} & \textbf{UTMOS↑} & \textbf{SIM↑}  & \textbf{WER↓} & \textbf{UTMOS ↑} & \textbf{CER↓} & \textbf{UTMOS ↑} \\
\midrule
UniAudio~\cite{DBLP:journals/corr/abs-2503-10522}                     & 7.2 & 3.46 & 0.40 & -     & -    & -    & 20.2 & 3.26 &-      &- \\
Mini-CPM-O-2.6~\cite{yao2024minicpm}            & 3.4 & 3.49 & 0.36 & 13.0 & 2.94 & 0.47 & 11.1 & 3.76 & 13.1 & 3.30 \\
Qwen2.5-Omni~\cite{DBLP:journals/corr/abs-2503-20215}                 & 2.1 & 4.16 & -    & 1.6 & 3.28 & -    & 7.6 &  4.19 & 2.5 & 3.38 \\
Step-audio~\cite{DBLP:journals/corr/abs-2502-11946}                & 2.2 & 3.84 & 0.52 & 1.0 & 3.23 & 0.62 & 5.0 &  4.37 & 2.7 & 3.69 \\
Step-audio 2 mini~\cite{DBLP:journals/corr/abs-2507-16632}        & 1.6 & 4.22 & 0.47 & 1.6 & 3.40 & 0.63 & \bt{3.5}  & \bt{4.35}  & 3.2  &\bt{4.00}   \\
Higgs audio V2~\cite{higgsaudio2025}            & \bt{1.0} & 4.00 & \bt{0.67} & \bt{0.8} & 3.27 & \bt{0.73} & 3.6 & 4.26 & 5.9 & 3.89 \\
MiMo~\cite{coreteam2025mimoaudio}            & 4.6 & 3.06 & - & 1.0 & 2.35 & - & 7.3 & 2.83 & 6.9 & 2.32 \\
\cmidrule(lr){1-11}
\textit{Unify-Baseline} & 2.5 & 3.67 & 0.47 & 2.0 & 3.29 & 0.57 & 10.8 & 3.97 & 4.2 & 3.45 \\
\textit{UniMoE-Audio}   & 1.9 & \bt{4.36} & 0.56 & \bt{0.8} & \bt{3.73} & 0.65 & 4.4 & 4.23 & \bt{1.6} & 3.86 \\
\bottomrule
\end{tabular}
\end{table*}

\begin{table*}[ht]
\centering
\caption{Performance on text-to-music and video-to-music generation benchmarks. The best performance for each metric is highlighted in \textbf{bold}. PC, PQ, and CE measure the aesthetic quality (higher is better). CLAP and CLaMP3 measure semantic alignment between the description and generated music (higher is better). KL and FAD assess audio quality against reference tracks (lower is better), while IS assess audio diversity (higher is better). 
UniMoE-Audio demonstrates superior performance in aesthetic quality, while remaining highly competitive in semantic alignment and audio quality.}
\label{tab:Music_main}
\begin{tabular}{l|l|l|cccccccc}
\toprule
\textbf{Dataset} & \textbf{Method} & \textbf{Task} & \textbf{PC↑} & \textbf{PQ↑} & \textbf{CE↑} & \textbf{CLAP↑} & \textbf{KL↓} & \textbf{CLaMP3↑} & \textbf{IS↑ } &\textbf{FAD↓} \\
\midrule
\multirow{8}{*}{MusicCap} 
& YuE~\cite{DBLP:journals/corr/abs-2503-08638}                           & T2M & 3.45 & 7.25 & 5.84 & 0.18 & 2.12 & 0.09 & 2.09 & 9.02 \\
& Stable Audio Open 1.0~\cite{DBLP:conf/icassp/EvansPCZTP25}         & T2M & 3.70 & 7.29 & 6.02 & \bt{0.30} & 1.44 & 0.11 & 2.74 & 3.72 \\
& AudioX~\cite{DBLP:journals/corr/abs-2503-10522}                        & T2M & 5.00 & 6.67 & 6.14 & 0.25 & \bt{1.20} & \bt{0.12} & \bt{3.02} & \bt{1.64} \\
& MusicGen~\cite{DBLP:conf/nips/CopetKGRKSAD23}                         & T2M & 4.78 & 7.37 & 6.57 & 0.26 & 1.21 & 0.10 & 1.68 & 7.02 \\
& MUMU-LLAMA~\cite{DBLP:journals/corr/abs-2412-06660}                    & T2M & 5.15 & 7.71 & 6.87 & 0.20 & 1.27 & 0.10 & 1.44 & 8.57 \\
\cmidrule(lr){2-11}
& \textit{Unify-Baseline}   & T2M & 5.66 & 6.48 & 5.30 & 0.14 & 1.57 & 0.07 & 1.57 & 9.64 \\
& \textit{UniMoE-Audio}     & T2M & \bt{6.00} & \bt{7.77} & \bt{7.34} & 0.29 & 1.39 & \bt{0.12} & 1.93 & 6.43 \\
\midrule
\multirow{8}{*}{V2M-bench} 
& YuE~\cite{DBLP:journals/corr/abs-2503-08638}                           & T2M & 3.78 & 7.25 & 6.01 & 0.15 & 1.27 & 0.13 & 1.79 & 4.29 \\
& Stable Audio Open 1.0~\cite{DBLP:conf/icassp/EvansPCZTP25}         & T2M & 3.41 & 7.46 & 5.69 & \bt{0.34} & 1.91 & 0.16 & 3.13 & 2.94 \\
& AudioX~\cite{DBLP:journals/corr/abs-2503-10522}                        & T2M & 4.60 & 7.30 & 6.06 & 0.30 & 2.12 & 0.11 & \bt{3.64} & 4.26 \\
& MusicGen~\cite{DBLP:conf/nips/CopetKGRKSAD23}                         & T2M & 4.64 & 7.37 & 6.24 & 0.28 & 1.27 & 0.15 & 1.70 & 3.39 \\
& MUMU-LLAMA~\cite{DBLP:journals/corr/abs-2412-06660}                    & T2M & 5.19 & 7.73 & 6.75 & 0.17 & \bt{0.92} & 0.13 & 1.42 & \bt{2.54} \\
\cmidrule(lr){2-11}
& \textit{Unify-Baseline}   & T2M & 5.71 & 5.68 & 4.33 & 0.23 & 1.89 & 0.15 & 1.83 & 3.27 \\
& \textit{UniMoE-Audio}     & T2M & \bt{5.75} & \bt{7.58} & \bt{6.85} & 0.31 & 1.06 & \bt{0.19} & 2.17 & 3.11 \\
\midrule
\multirow{3}{*}{V2M-bench} 
& AudioX~\cite{DBLP:journals/corr/abs-2503-10522}                        & V2M & 4.44 & 7.44 & 6.06 & - & 1.84 & - & 3.14 & 2.94 \\
\cmidrule(lr){2-11}
& \textit{Unify-Baseline}     & V2M & 4.61 & 5.50 & 4.29 & - & 2.01 & - & 1.74 & 3.24 \\
& \textit{UniMoE-Audio}     & V2M & \bt{5.88} & \bt{7.62} & \bt{6.96} & - & \bt{1.69} & - & \bt{3.31} & \bt{2.89} \\
\bottomrule
\end{tabular}
\end{table*}

We conducted a comprehensive evaluation of UniMoE-Audio against state-of-the-art specialized models and strong baselines across a variety of speech and music generation tasks. As detailed in Table~\ref{tab:TTS_main} and Table~\ref{tab:Music_main}, our results demonstrate that UniMoE-Audio model can achieve competitive or even superior performance in both domains, effectively overcoming the typical trade-offs associated with joint multi-task training.

\noindent\textbf{Takeaway 1: UniMoE-Audio achieves strong performance in speech synthesis with remarkable data efficiency}. 
As shown in Table~\ref{tab:TTS_main}, UniMoE-Audio demonstrates exceptional capabilities in speech synthesis. For example, on the SeedTTS-EN benchmark, it achieves a new state-of-the-art in perceptual quality with a UTMOS of 4.36, while also delivering highly competitive intelligibility (WER 1.9). This strong performance is also observed in other datasets. Notably, this performance is achieved using only 280K hours of speech data, rivaling or even surpassing dedicated models like Higgs audio V2 and Step-audio 2 mini, which were trained on 10M hours and 8M hours speech data. This highlights the remarkable data efficiency and strong learning capability endowed by our unified architecture and training curriculum. However, we also observe that the performance of speaker similarity remains inferior to the state-of-the-art model, which may be attributable to insufficient data scale.

\noindent\textbf{Takeaway 2: UniMoE-Audio excels in generating aesthetically superior music with strong semantic relevance}. 
In the domain of music generation (Table~\ref{tab:Music_main}), UniMoE-Audio consistently prioritizes and achieves superior aesthetic quality. Across both T2M and V2M tasks, our model obtains the highest scores in all aesthetic metrics (PC, PQ, CE), indicating its strength in producing richer, more enjoyable musical content. While its reference-similarity-based audio quality scores (i.e. FAD and KL) are inferior, we posit that this reflects our model's strength in creative generation rather than mere imitation of reference tracks, which explores a broader and more diverse acoustic space. Furthermore, the model attains strong semantic alignment with textual prompts, as evidenced by high CLAP and CLaMP3 scores. This combination of superior aesthetic quality and precise semantic alignment demonstrates UniMoE-Audio’s capability as a powerful and versatile music generation system.

\noindent\textbf{Takeaway 3: The MoE architecture is critical for mitigating task conflict and enabling multi-domain excellence}. 
A direct comparison between UniMoE-Audio (MoE) and Unify-Baseline (Dense) provides  strong empirical evidence supporting our architectural choice. Across both speech and music domains, the dynamic-capacity MoE consistently and significantly outperforms the dense baseline, despite the similar model size. This stark performance gap demonstrates that naive joint training leads to catastrophic interference, whereas our dynamic-capacity MoE architecture, by dynamically activating specialized experts, effectively resolves this conflict and unlocks high performance in both domains.

\noindent\textbf{Takeaway 4: Our training approach effectively mitigates the inherent data imbalance in multi-task learning}. 
The Unify-Baseline model serves as a stark illustration of the catastrophic forgetting induced by data imbalance in naive joint training. While its performance on the data-dominant Mandarin TTS task (comprising about 40\% of the data) remains reasonable, its ability to generate coherent music is severely compromised, as evidenced by its poor music generation performance. In stark contrast, UniMoE-Audio demonstrates robust performance even on the most resource-limited task, Video-to-Music (V2M), which constitutes merely 5\% of the training data. This success is directly attributable to our methodology. By first training "proto-experts" on individual tasks, we pre-assign their specialized roles. The subsequent MoE integration then allows the model to dynamically route inputs to the relevant experts, effectively preventing the knowledge of data-scarce tasks from being overwritten during joint training. This demonstrates that our approach effectively mitigates the typical pitfalls of naive joint training, preserving high-quality generation capabilities across all supported domains, irrespective of their data representation.

\section{Discussion}

\begin{figure}[htbp!]
\centering
    {\includegraphics[width=0.9\columnwidth]{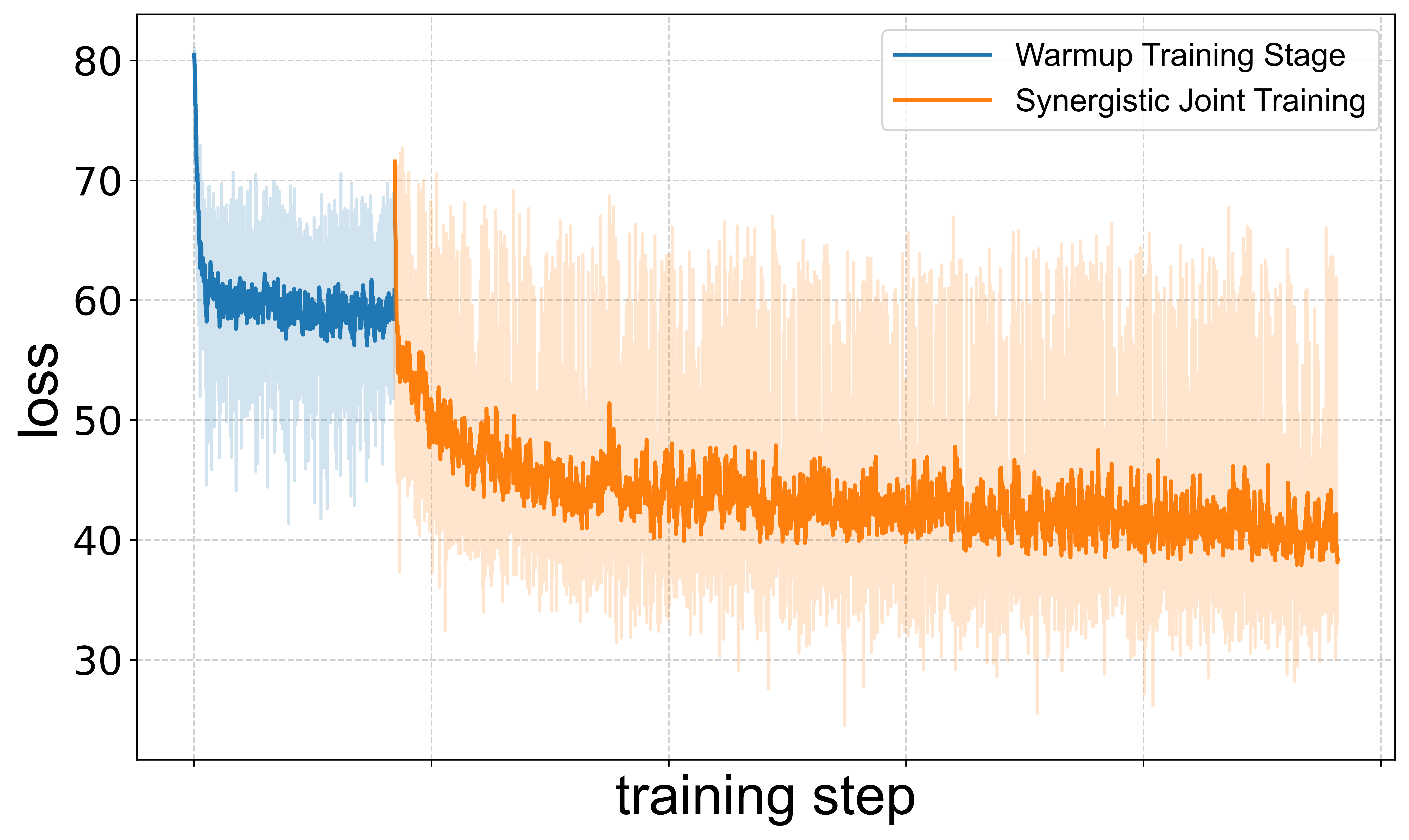}}
    {\includegraphics[width=0.9\columnwidth]{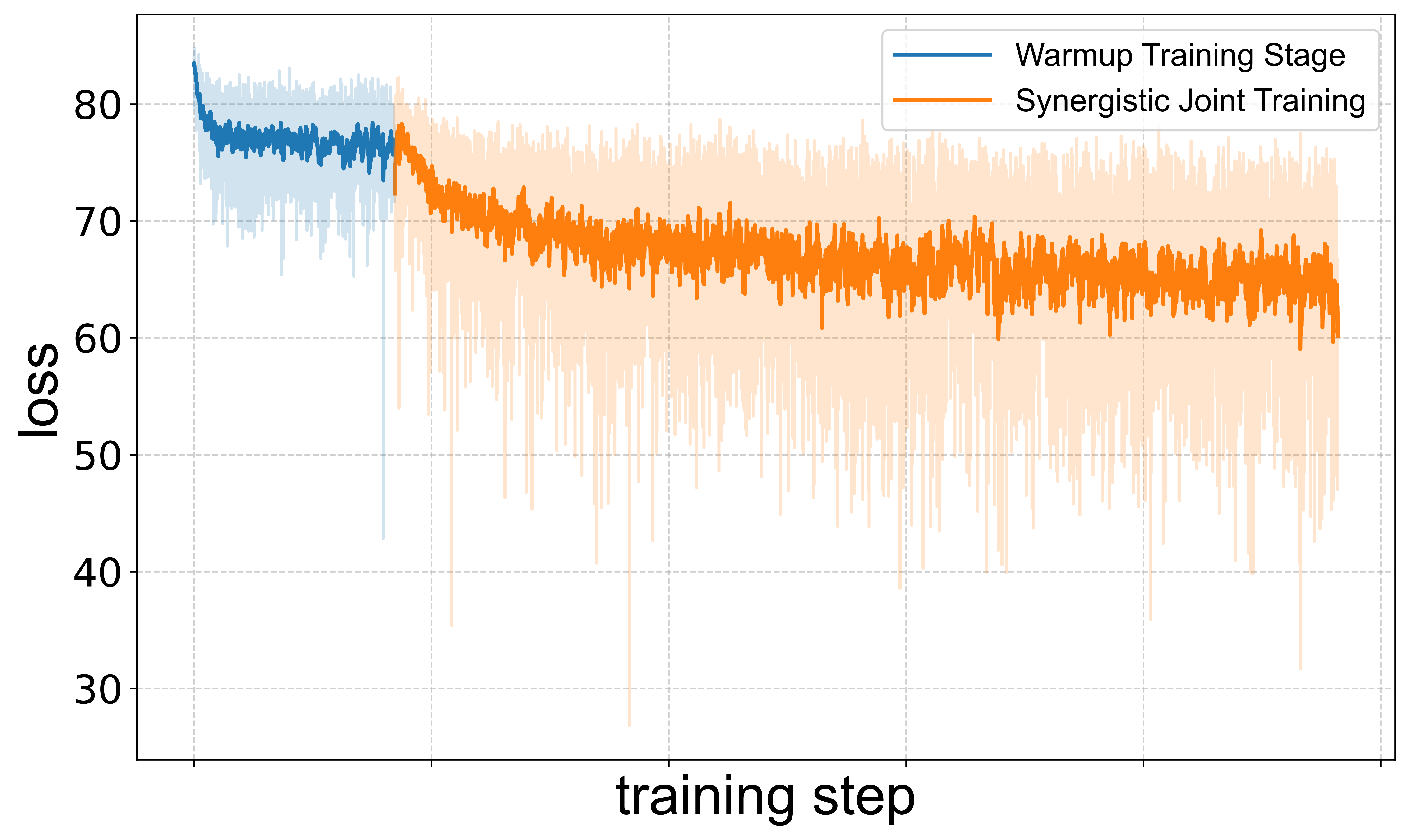}}
    \caption{Training loss for the speech generation task (top) and music generation task (bottom). The plots show the transition from the Warmup Training Stage (blue) to the Synergistic Joint Training Stage (orange). The solid line represents the moving average of the loss.}
    \label{fig:losses}
\end{figure}

In this section, we delve into a deeper analysis of UniMoE-Audio's training dynamics and internal mechanics. We first examine the training loss curves to gain insights into our three-stage curriculum (\S\ref{sec:loss}). Subsequently, we analyze the expert utilization patterns to understand how the model allocates its capacity across different layers (\S\ref{sec:topk}). Finally, we analyze the distribution of expert loading across speech and music generation tasks  (\S\ref{sec:expert}).

\subsection{Training Loss Analysis}
\label{sec:loss}

Figure~\ref{fig:losses} illustrates the training loss dynamic for speech (top) and music (bottom) generation, separated into the warmup and synergistic joint training stages. These curves provide several key insights into our training curriculum:

\noindent\textbf{Warmup Stage is Essential for Stable Router Calibration.} The loss reduction magnitude during the warmup phase is comparable to that of the subsequent joint training phase, underscores its critical role. This confirms that calibrating the routing mechanism is a non-trivial optimization problem. Our staged approach effectively decouples this from expert optimization, allowing the router to learn stable expert dispatch patterns before full-model training, thus preventing initial instability from corrupting the pre-trained experts.

\noindent\textbf{Staged Training Enhances Overall Stability.} The joint training phase exhibits higher loss volatility compared to the smoother warmup phase across both tasks. This suggests that end-to-end training of the full MoE is inherently less stable. By pre-stabilizing the routing logic during the warmup, our curriculum mitigates the risk of suboptimal performance and ensures a more robust convergence path during the final joint training stage.

\noindent\textbf{Loss Disparity Reflects Intrinsic Task Complexity.} The music generation task consistently shows a higher loss than the speech task (converging near 60 vs. 40). This empirically validates our hypothesis that music, with its complex structures, is an intrinsically more difficult task to model. This difficulty gap highlights the necessity of our MoE architecture and staged curriculum, which prevent the "easier" speech task from dominating the learning process, a problem often seen in naive joint training.

\subsection{Analysis of Dynamic Expert Allocation}
\label{sec:topk}

\begin{figure}
\centering
\includegraphics[width=\linewidth]{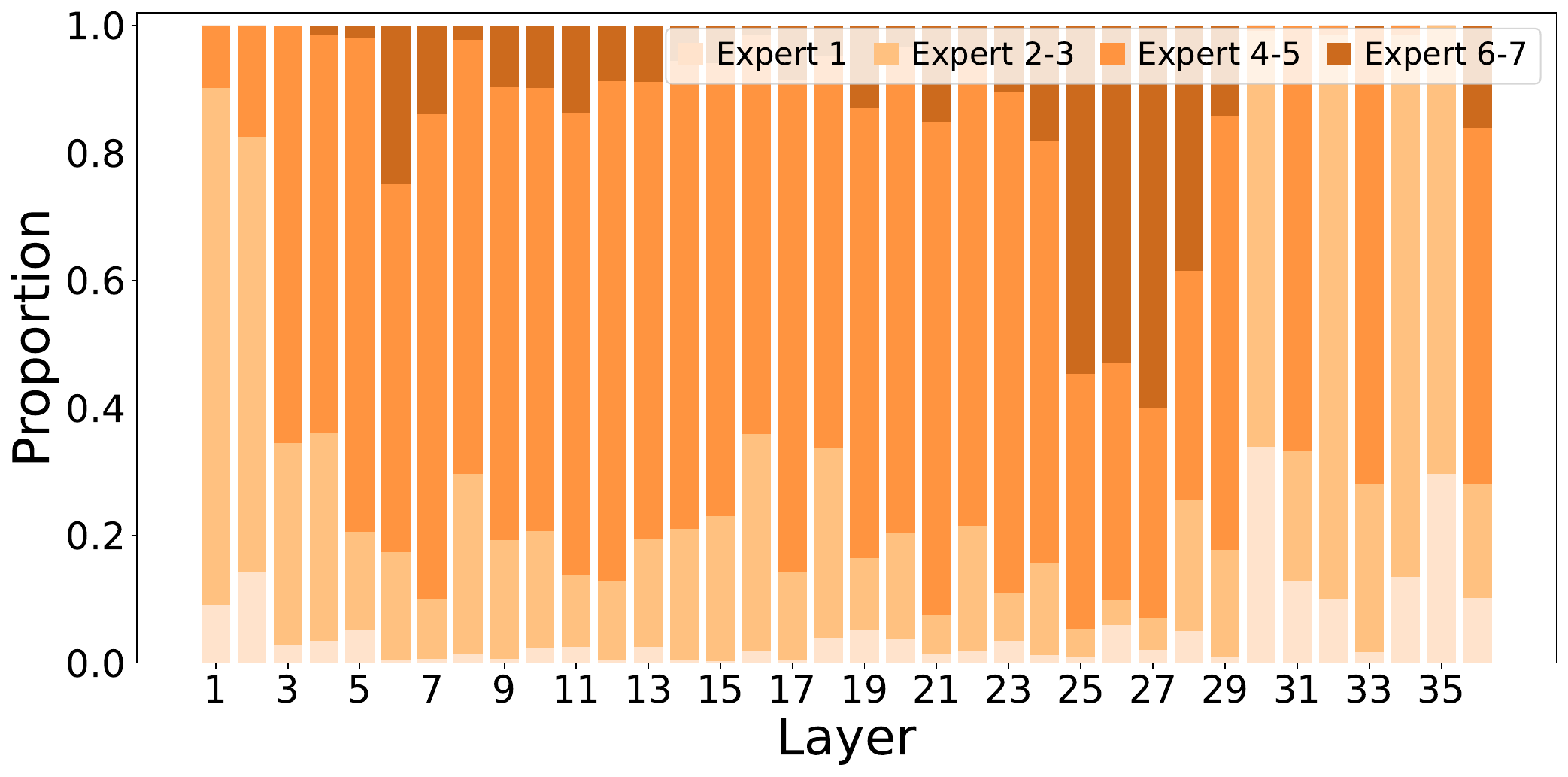}
\caption{Visualization of the dynamic computational budget allocated by our Top-P routing mechanism. The figure illustrates the proportion of tokens activating a varying number of experts at each layer, revealing a "rise-and-fall" pattern where more computational resources are adaptively assigned to the middle layers. For clarity, counts of activated experts are grouped into bins (e.g., "Expert 2-3" represents tokens activating either 2 or 3 experts).}
\label{fig:topk_visualization}
\end{figure}

\begin{figure*}[htbp]
    \centering 
    {\includegraphics[width=0.19\textwidth]{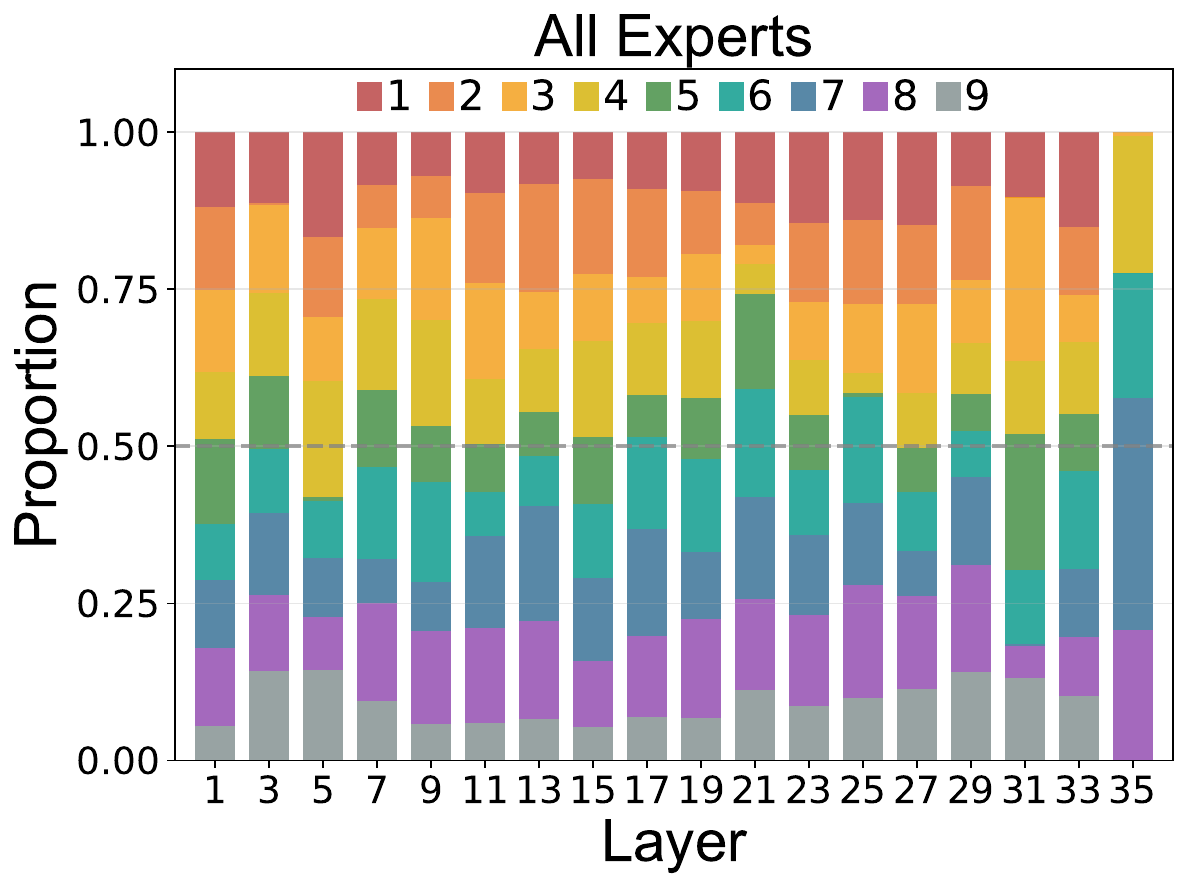}}
    {\includegraphics[width=0.19\textwidth]{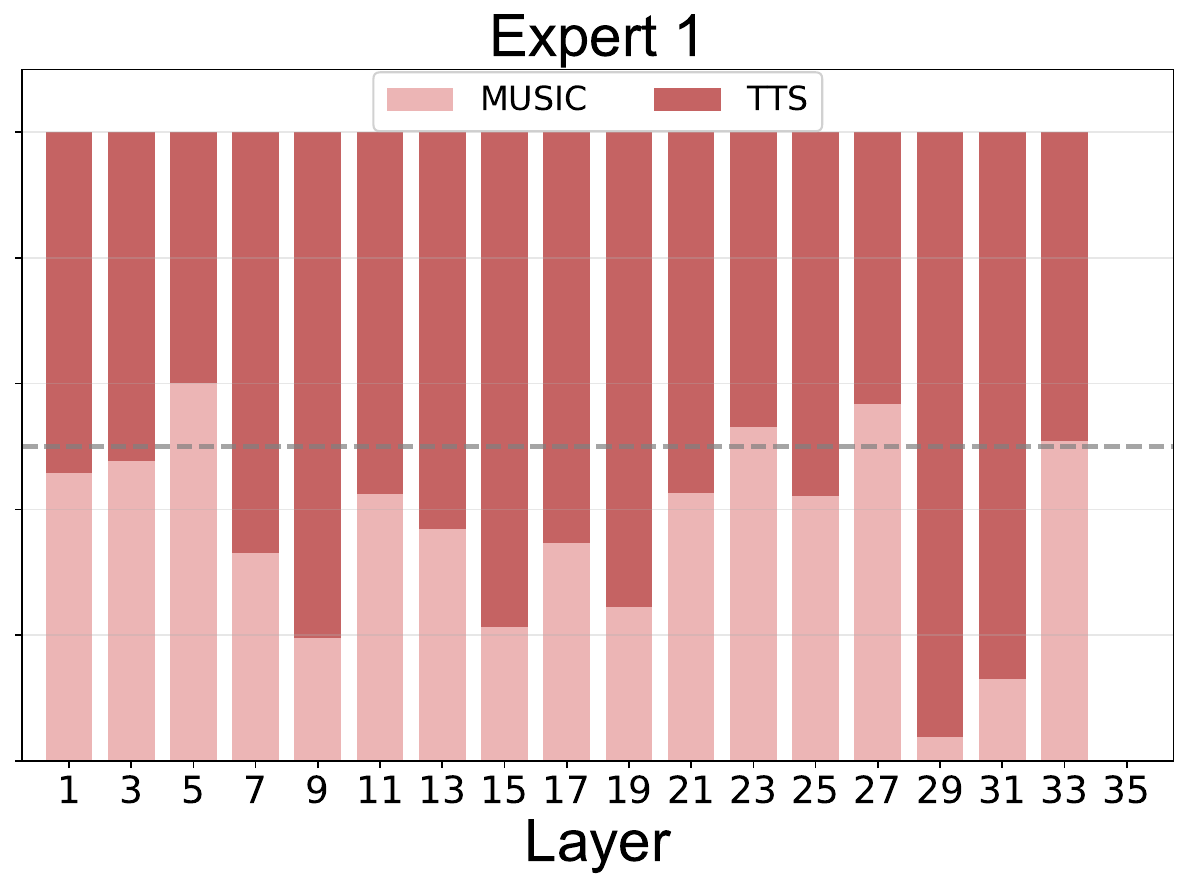}}
    {\includegraphics[width=0.19\textwidth]{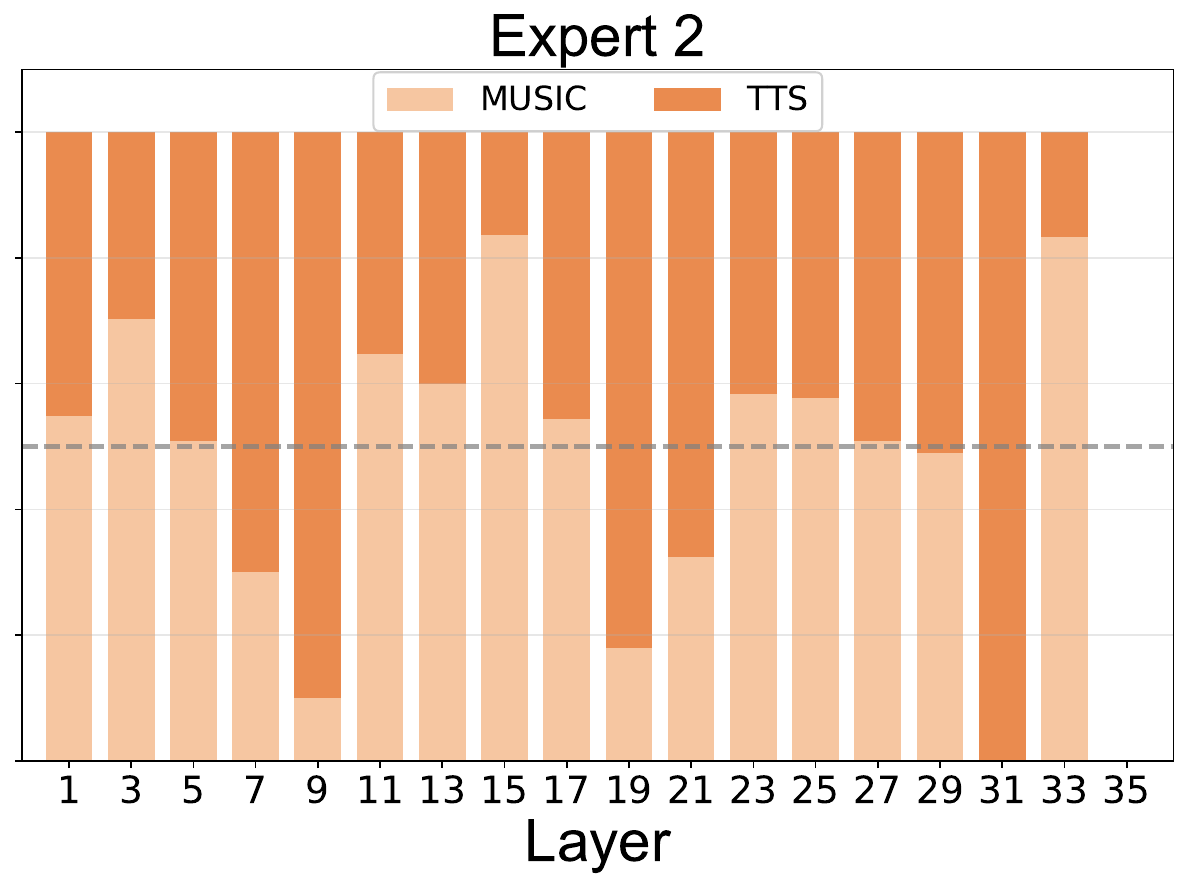}}
    {\includegraphics[width=0.19\textwidth]{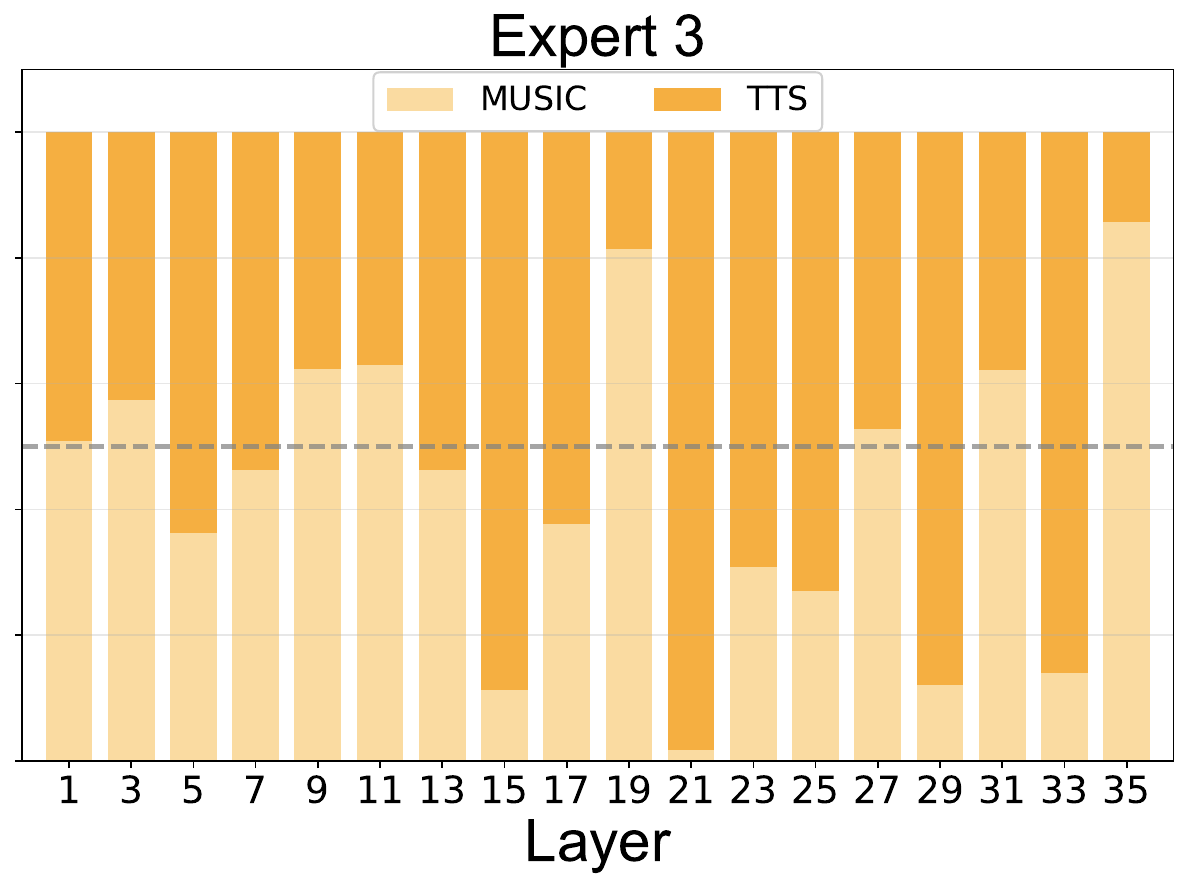}}
    {\includegraphics[width=0.19\textwidth]{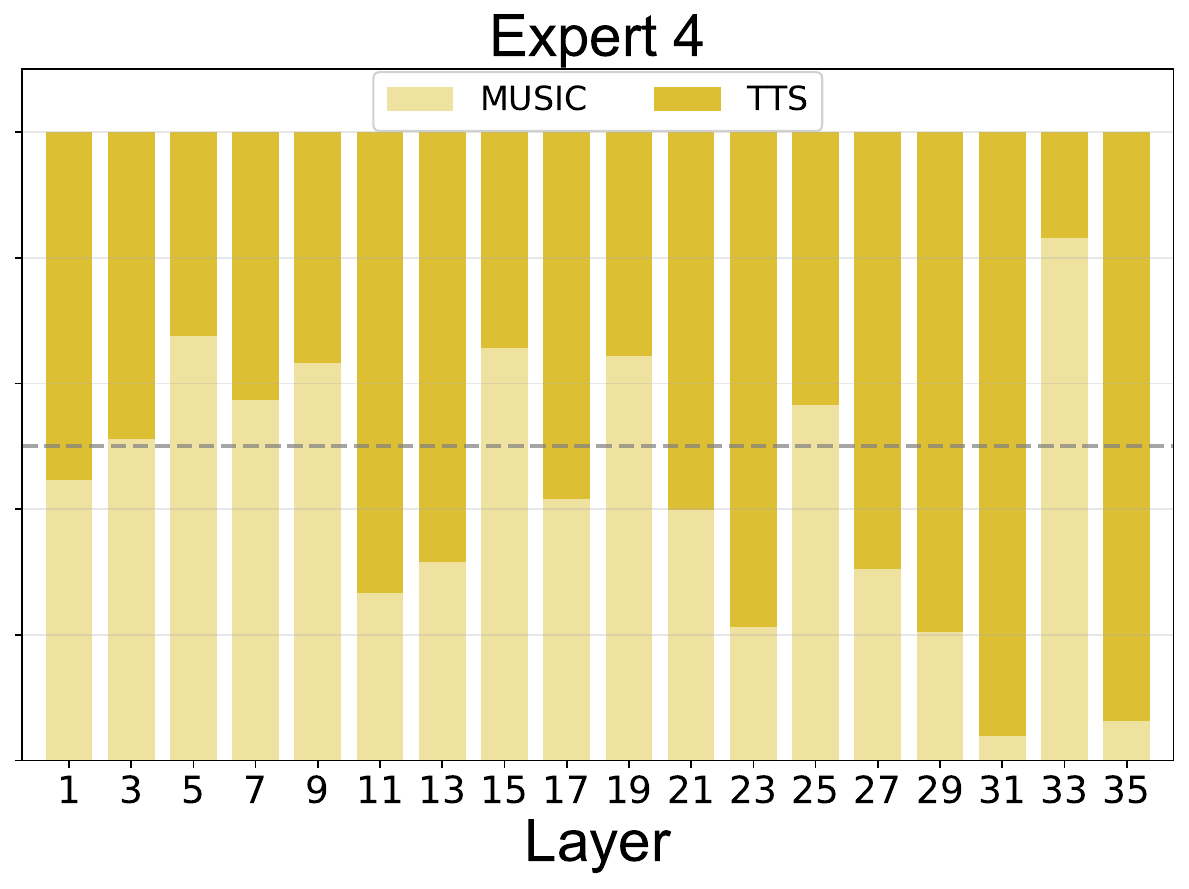}}
    \vspace{0.5cm} 
    {\includegraphics[width=0.19\textwidth]{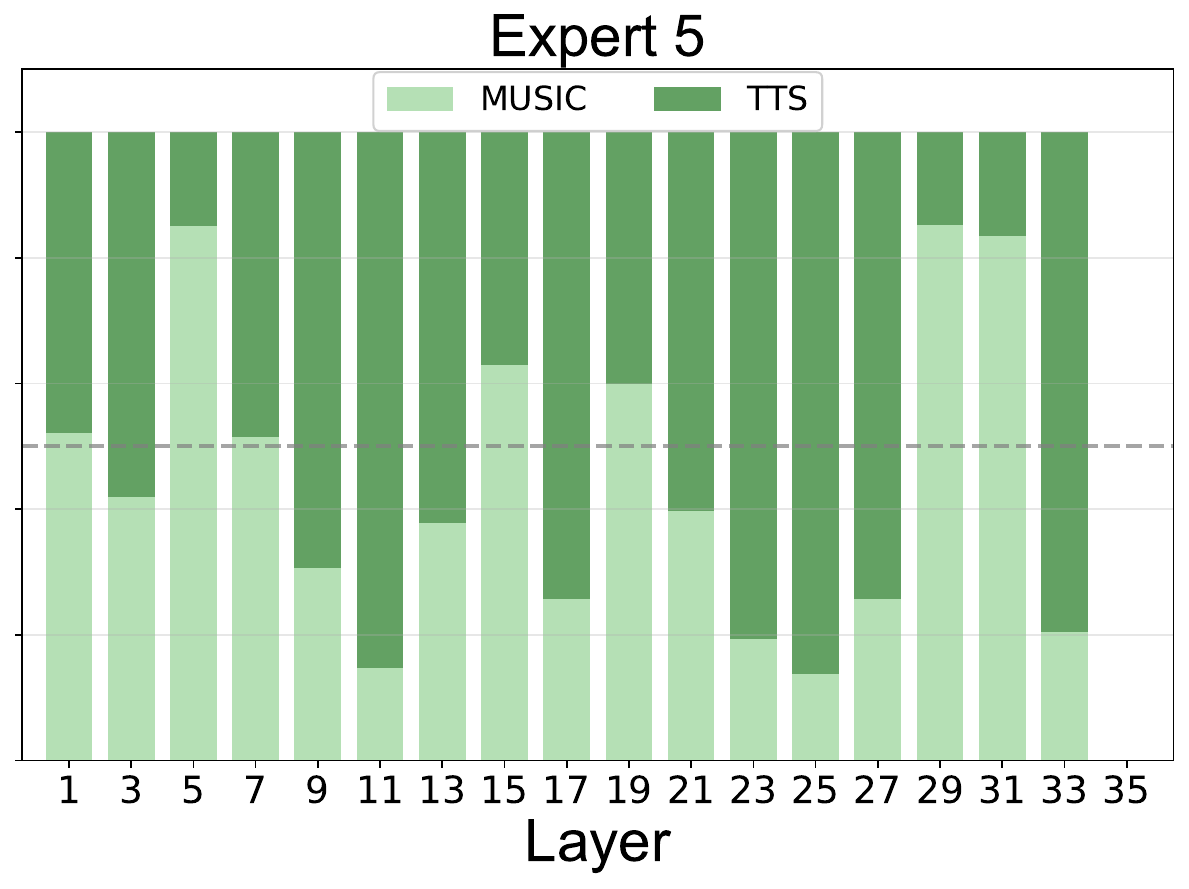}}
    {\includegraphics[width=0.19\textwidth]{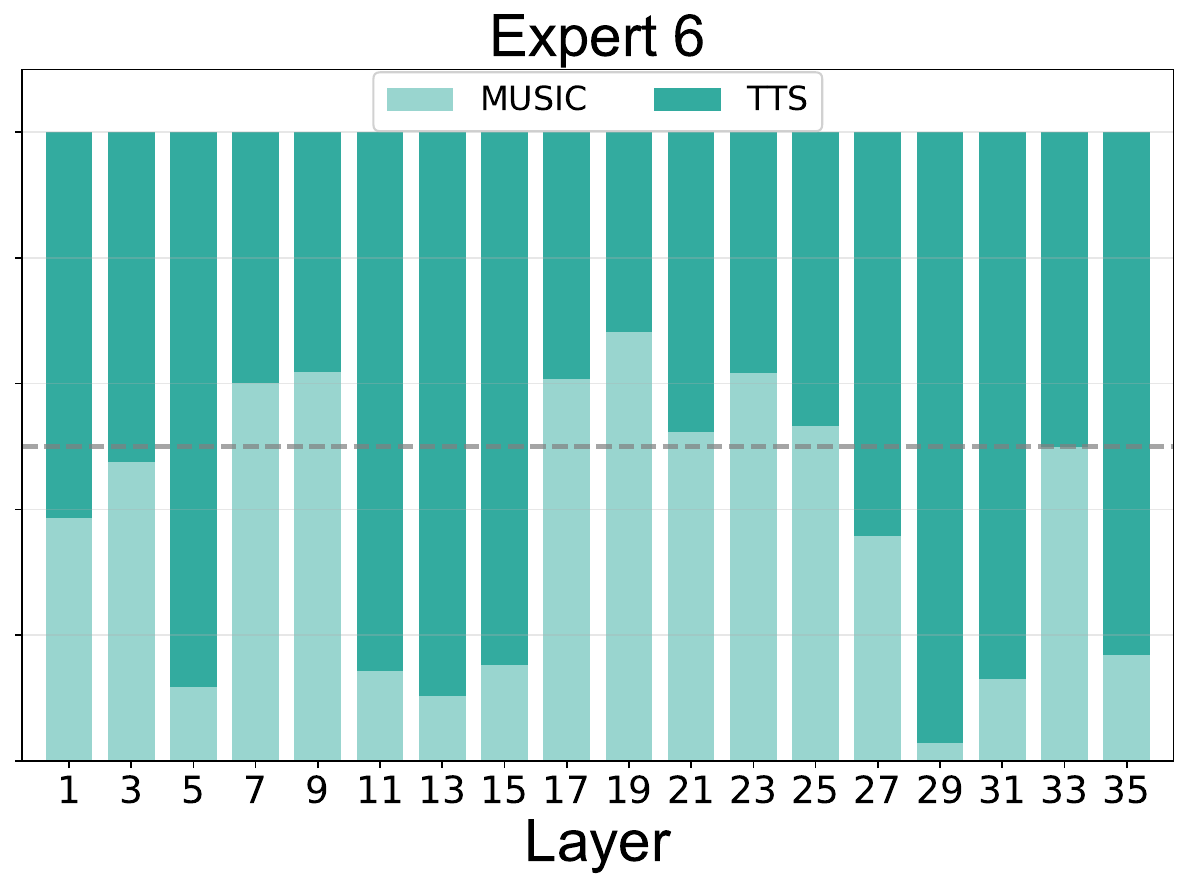}}
    {\includegraphics[width=0.19\textwidth]{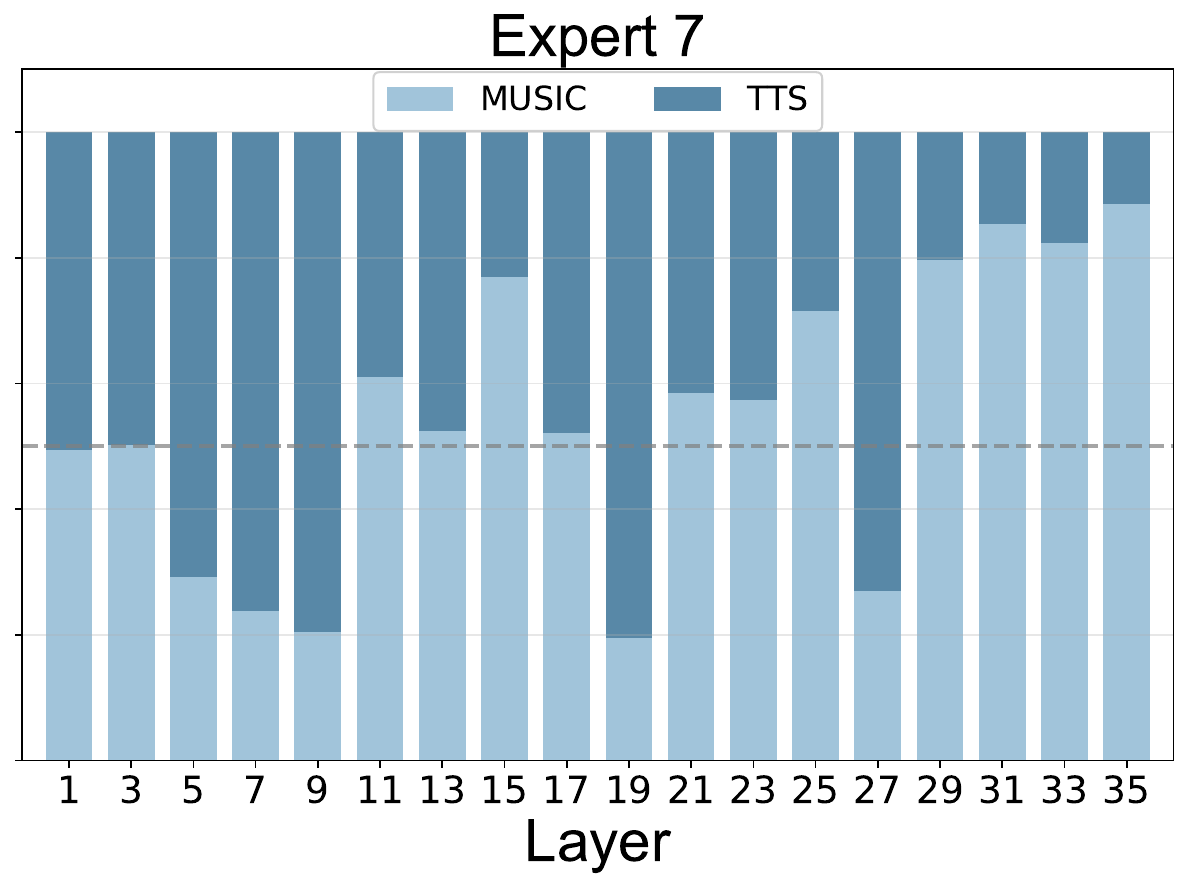}}
    {\includegraphics[width=0.19\textwidth]{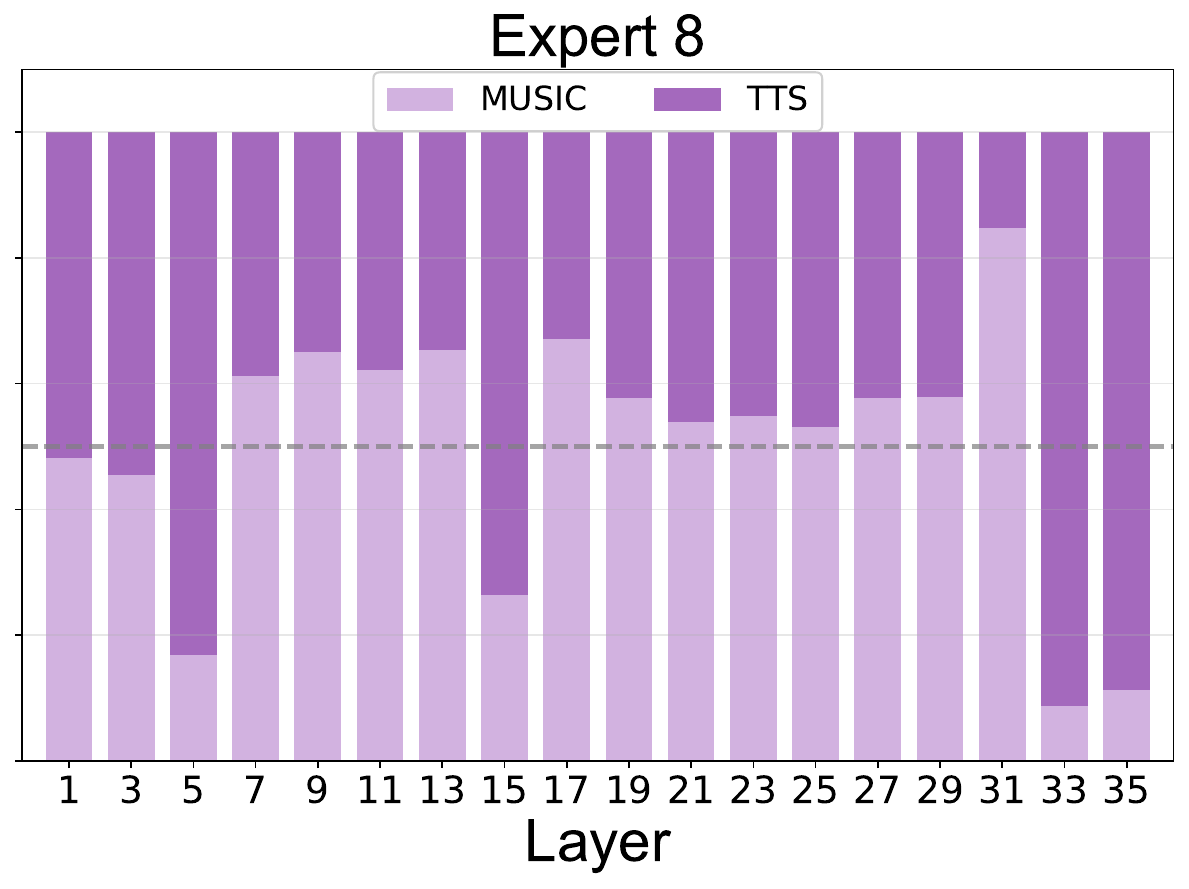}}
    {\includegraphics[width=0.19\textwidth]{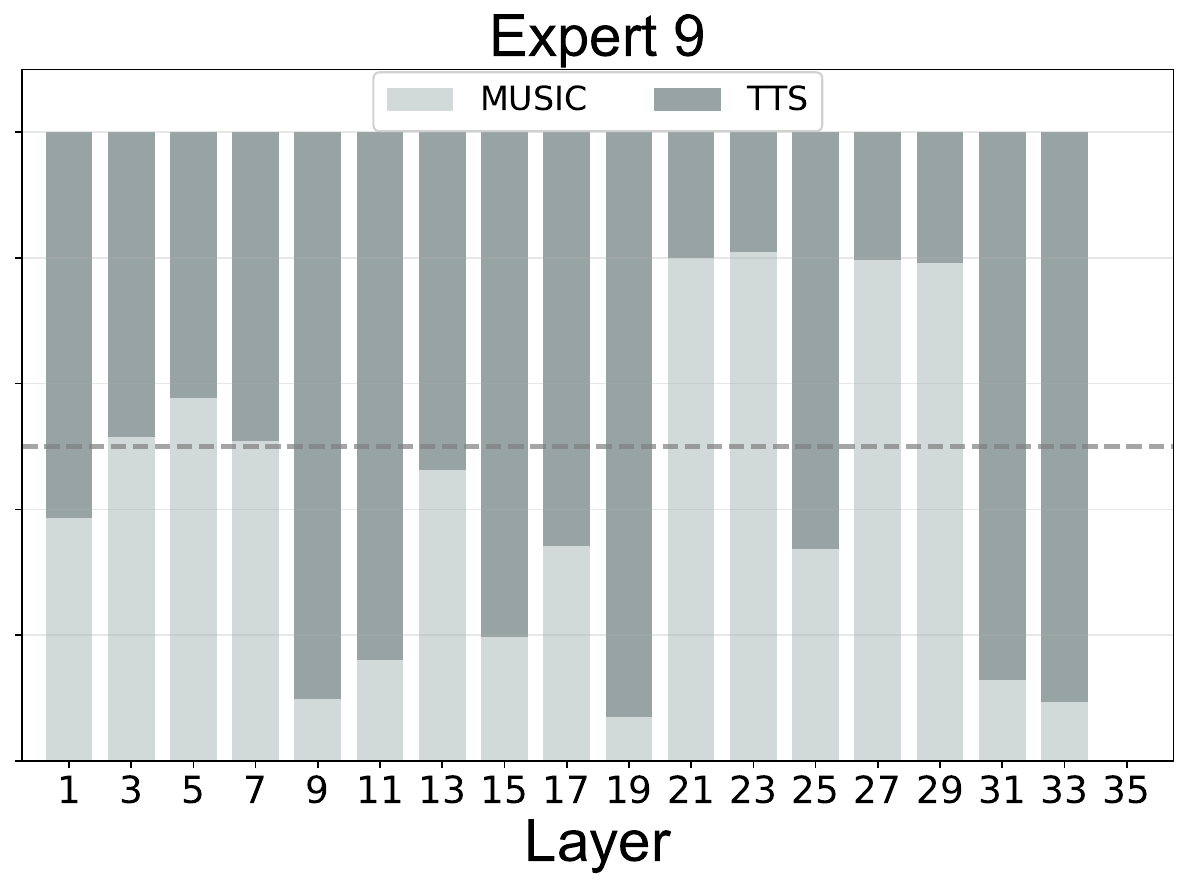}}
    \caption{Analysis of expert routing dynamics in UniMoE-Audio across transformer layers. The top-left "All Experts" plot illustrates the routing frequency for each of the eight routed experts (E1-E8, colored) and the null expert (E9, gray). The subsequent nine plots provide a granular breakdown for each expert, showing the proportion of tokens routed from the Music (lighter shade) versus the TTS (darker shade) task.}
    \label{fig:expert_roll}
\end{figure*}

To investigate the operational dynamics of our Top-P routing strategy, we analyze the distribution of the number of activated experts per token across different layers, as shown in Figure~\ref{fig:topk_visualization}. The visualization reveals a clear pattern of hierarchical computational demand. In the initial layers (e.g., layers 0-3), most tokens are routed to a smaller number of experts (typically 1-3). This likely corresponds to low-level feature extraction. As information propagates to the middle layers (e.g., layers 4-13), it allocates a larger computational budget, with the majority of tokens activating 4-5 experts. This allocation peaks around layer 12, indicating that the model concentrates its most intensive computations here for complex feature abstraction and cross-modal fusion. Subsequently, in the final, deeper layers (14-17), the trend reverses, and the allocated budget decreases again, likely focusing on integrating features for final output generation.

Crucially, this non-uniform, layer-wise allocation pattern highlights a core advantage of Top-P routing over conventional Top-K. A common Top-K strategy would enforce a fixed computational budget at every layer, irrespective of the layer's function. In contrast, our model learns to dynamically tailor its capacity, assigning more resources to the middle layers. Furthermore, even within a single layer, the distribution is not monolithic; the model adaptively assigns a larger budget to "hard" tokens while conserving resources on "easy" ones. This inherent flexibility validates the efficacy of Top-P routing in creating a more efficient and intelligent architecture that allocates its computational power precisely where it is needed most.

\subsection{Expert Routing Visualization}
\label{sec:expert}

To delve into the expert loading distribution of UniMoE-Audio, we visualize the expert routing statistics in Figure~\ref{fig:expert_roll}. The figure provides a comprehensive overview, showing the overall expert utilization (top-left) and a detailed breakdown of task preference (Music vs. Speech) for each of the eight routed experts (E1-E8) and the null expert (E9). The analysis reveals several key findings.

\noindent\textbf{Effective Load Balancing Prevents Expert Collapse.} The "All Experts" subplot shows a remarkably balanced workload across all layers. No single routed expert is either over-utilized or ignored, and the null expert (E9) is also consistently engaged. This demonstrates that our training approach successfully prevents expert collapse—a common failure mode in MoE training where the router shows strong preference for certain experts. This balanced utilization confirms that all experts are actively contributing to the model's computation.

\noindent\textbf{Experts Exhibit Clear and Consistent Task Specialization.} The individual plots for Experts 1 through 8 provide striking evidence of learned task specialization. A clear division of labor is visible: Experts 1-4 consistently show a strong preference for Speech tokens, while Experts 5-8 are overwhelmingly activated by Music tokens. For instance, across most layers, Expert 1 (red) is almost exclusively chosen for TTS, whereas Expert 5 (green) is predominantly chosen for Music. This strong, persistent specialization directly validates our training strategy. Initializing the model with pre-trained "proto-experts" successfully instills domain-specific knowledge, and the subsequent training preserves these roles, allowing the model to route tasks to the most qualified specialist.

\noindent\textbf{Hierarchical Processing Emerges from Shallow to Deep Layers.}
While specialization is strong overall, we find that experts of initial layers (e.g., 1-5) show a more mixed activation between TTS and Music compared to the deeper layers. For example, in Expert 2 and Expert 6, the proportion of the non-preferred task is visibly higher in the first few layers. This suggests an emergent hierarchical processing scheme: shallower layers likely handle more universal, low-level features common to both speech and music (e.g., basic frequencies), while deeper layers focus on processing more abstract, domain-specific information, such as phonetics for TTS or harmony for Music.

\noindent\textbf{The Role of the Null Expert in Adaptive Computation} The behavior of the null expert (E9) provides a profound insight into the model's learned efficiency. While the "All Experts" plot shows it handles a substantial workload, the dedicated "Expert 9" plot reveals a dynamic, layer-dependent preference. In shallower layers, it prunes simple tokens from both tasks equally. However, in the deeper layers (25-32), it is overwhelmingly activated by speech tokens. This strongly suggests that once high-level features are formed, the model identifies the TTS task as computationally simpler and learns to skip redundant computations for it. This not only validates the null expert as a mechanism for learned efficiency but also provides empirical evidence that our model dynamically understands the varying complexity of each task across its depth.

\section{Conclusion}

In this paper, we addressed the long-standing challenge of unifying speech and music generation, a task hindered by task conflict and data imbalance. We introduced UniMoE-Audio that leverages a dynamic-capacity Mixture-of-Experts architecture to mitigate task conflict, in conjunction with a data-aware, three-stage training curriculum to overcome data imbalance. Experiments across diverse benchmarks show that UniMoE-Audio not only matches or surpasses strong domain-specific baselines, but also enables synergistic learning across audio domains—effectively avoiding the performance degradation observed in naive joint training. Our work provides a robust blueprint for building unified generative audio models, with future directions include the incorporation of a broader range of audio types and the optimization of MoE architecture.

\bibliographystyle{IEEEtran}
\bibliography{custom}

\appendices

\vfill

\end{document}